**Absence of detectable arsenate in DNA from arsenate-grown GFAJ-1 cells.**

Marshall Louis Reaves[1,2], Sunita Sinha[3], Joshua D. Rabinowitz[1,4], Leonid Kruglyak[1,5,6], and Rosemary J. Redfield[3†].

[1]Lewis Sigler Institute for Integrative Genomics, Princeton University, Princeton, NJ, USA. [2]Department of Molecular Biology, Princeton University, Princeton, NJ, USA. [3] Department of Zoology, University of British Columbia, Vancouver, BC, Canada. [4]Department of Chemistry, Princeton University, Princeton, NJ, USA. [5]Howard Hughes Medical Institute, Lewis-Sigler Institute for Integrative Genomics. [6]Department of Ecology and Evolutionary Biology, Princeton University, Princeton, NJ, USA.

[†] Corresponding author: redfield@zoology.ubc.ca

**Abstract**

A strain of *Halomonas* bacteria, GFAJ-1, has been reported to be able to use arsenate as a nutrient when phosphate is limiting, and to specifically incorporate arsenic into its DNA in place of phosphorus. However, we have found that arsenate does not contribute to growth of GFAJ-1 when phosphate is limiting and that DNA purified from cells grown with limiting phosphate and abundant arsenate does not exhibit the spontaneous hydrolysis expected of arsenate ester bonds. Furthermore, mass spectrometry showed that this DNA contains only trace amounts of free arsenate and no detectable covalently bound arsenate.

Wolfe-Simon et al. isolated strain GFAJ-1 from the arsenic-rich sediments of California's Mono Lake by its ability to grow through multiple subculturings in artificial Mono Lake medium AML60 that lacked added phosphate but had high concentrations of arsenate (+As/-P condition) (*1*). Because GFAJ-1 grew in -P medium only when arsenate was provided, and because significant amounts of arsenate were detected in subcellular fractions, growth was attributed to the use of arsenate in place of phosphate. However, the basal level of phosphate contaminating the -P medium was reported to be 3-4 μM (*1*), which previous studies of low-phosphate microbial communities suggest is sufficient to support moderate growth (*2*). GFAJ-1 grew well on medium supplemented with ample phosphate but no arsenate (1500 μM $PO_4$, +P/-As condition), indicating that GFAJ-1 is not obligately arsenate-dependent.

Wolfe-Simon *et al*. (*1*) further reported that arsenic was incorporated into the DNA backbone of GFAJ-1 in place of phosphorus, with an estimated 4% replacement of P by As based on the As:P ratio measured in agarose gel slices containing DNA samples. This finding was surprising because arsenate is predicted to reduce rapidly to arsenite in physiological conditions (*3, 4*), and because arsenate esters in aqueous solution are known to be rapidly hydrolyzed (*5*). We have now tested this report by culturing GFAJ-1 cells supplied by the authors (*1*) and by analyzing highly purified DNA from phosphate-limited cells grown with and without arsenate.

Wolfe-Simon *et al*. reported that GFAJ-1 cells grew very slowly in AML60 medium (doubling time ~12 hours), and that, when phosphate was not added to the medium, cells failed to grow unless arsenate (40 mM) was provided (*1*). However, although we obtained strain GFAJ-1 from these authors, in our hands GFAJ-1 was unable to grow at all in AML60 medium containing the specified trace elements and vitamins, even with 1500 μM sodium phosphate added as specified in (*1*). We confirmed the strain's identity using RT-PCR and sequencing of 16S rRNA, using primers specified by Wolfe-Simon *et al*. (*1*); this gave a sequence identical to that reported for strain GFAJ-1. We then found that addition of small amounts of yeast extract, tryptone or individual amino acids to basal AML60 medium allowed growth, with doubling times of 90-180 minutes. Medium with 1 mM glutamate added was therefore used for subsequent experiments (*12)*.

With 1500 μM phosphate but no added arsenate (Wolfe-Simon *et al.*'s -As/+P condition), this medium produced ~ 2 x $10^8$ cells/ml, similar to the -As/+P yield obtained by Wolfe-Simon *et al.* (*1*). As expected, the growth yield depended on the level of phosphate supplementation (Fig. 1), with even unsupplemented medium allowing significant growth (~ 2 x $10^6$ cells/ml). As ICP-MS analysis showed that this medium contained only 0.5 μM contaminating phosphate, our supplementing with an additional 3.0 μM phosphate replicates Wolfe-Simon *et al.*'s '-P' culture condition. The growth analyses shown in Fig. 1 were performed in the absence of arsenate, and showed that GFAJ-1 does not require arsenate for growth in media with any level of phosphate.

The cause of the discrepancies between our growth results and those of Wolfe-Simon *et al.* is not clear. The arsenate dependence they observed may reflect the presence in their arsenate (purity and supplier unknown) of a contaminant that filled the same metabolic role as our glutamate supplement. Our +As and -As cultures grew to similar densities, and we never observed cases where +As cultures grew but -As cultures did not. The phosphate dependence we observed is also consistent with that expected from work on other species (*2*).

To investigate the possible incorporation of arsenate into the GFAJ-1 DNA backbone, we purified and analyzed DNA from GFAJ-1 cells grown in four differently supplemented versions of AML60 medium, matching those analyzed by Wolfe-Simon *et al. i.e.* -As/-P: no arsenate, 3.5 μM phosphate; +As/-P: 40 mM arsenate, 3.5 μM phosphate; -As/+P: no arsenate, 1500 μM phosphate; +As/+P: 40 mM arsenate, 1500 μM phosphate. Initial purification of DNA consisted of two preliminary organic extractions, precipitation from 70% ethanol, digestion with RNase and proteinase, two additional organic extractions, and a final ethanol precipitation (*12*). DNA was collected from 70% ethanol by spooling rather than centrifugation, since this reduces contamination with other substances insoluble in ethanol (*6*).

Wolfe-Simon *et al*. suggested that arsenate ester bonds in GFAJ-1 DNA might be protected from hydrolysis by intracellular proteins or compartmentalization of the DNA (*7*). We therefore tested whether purification exposed GFAJ-1 DNA to spontaneous hydrolysis. Gel analysis of DNA immediately after purification revealed fragments of > 30 kb, whether cells were grown with limiting or abundant phosphate, and with or without 40 mM arsenate (Fig. 2A). We also reexamined this DNA after two months of storage at 4 °C. All preparations showed very similar sized fragments of dsDNA and of ssDNA (Figs. 2B and 2C), with no evidence of hydrolysis. *Haemophilus influenzae* DNA served as a control for gel migration, indicating that GFAJ-1 DNA is not associated with hydrolysis-protecting proteins or other macromolecules that might have persisted through the purification. Unless arsenate-ester bonds are intrinsically stable in DNA, our analysis estimates a minimum separation between arsenates in the DNA backbone of at least 25 kb, three orders of magnitude below that estimated by Wolfe-Simon *et al*.

Arsenate in bonds that were stable to spontaneous hydrolysis should be detectable as free arsenate, arsenate-containing mononucleotides, or arsenate-containing di-nucleotides after enzymatic digestion of purified DNA. We therefore used liquid chromatography-mass spectrometry (LC-MS) to analyze GFAJ-1 DNA for arsenate after digestion with P1 and snake venom nucleases (*12*). Relevant molecular species were identified by negative mode, full scan, high-mass resolution LC-MS analysis (*12*). This method was used to analyze two independent replicate DNA preparations from cells grown in either +As/-P or –As/+P medium, and fractions from CsCl gradient analyses of these DNAs.

The initial DNA preparations of +As/-P DNAs contained some free arsenate anion ($H_2AsO_4^-$) (Table 1); similar to levels reported by Wolfe-Simon *et al*. (*1*). This arsenate was largely removed by three serial washes with distilled water; digested washed DNA contained arsenate at a level slightly higher than in the water blank (Fig. 3 and Table 1). Thus, we concluded that most of the arsenate we detected after preliminary DNA purification arose by contamination from the arsenate-rich (40 mM) growth medium.

Further analyses compared the nuclease-digested and washed fractions obtained from CsCl isopycnic density gradient centrifugation of the DNAs (Fig. 3) (*12*). The arsenate detection limit for these measurements was ~ 5 x $10^{-8}$ M (Table S1); a level that if present in the fractions with the most DNA would correspond to an As:P ratio of < 0.1%, 50-fold lower than the 4% ratio reported by Wolfe-Simon *et al*. Although traces of arsenate (or a contaminant of similar mass to arsenate) were found in several fractions of the CsCl gradient, the arsenate peak never exceeded the limit of detection, and a similar intensity signal at m/z of arsenate was observed in the water blank. There was no evidence that the arsenate trace co-migrated with the DNA. In contrast, normal phosphate-containing deoxynucleotides were observed in rough proportion to the abundance of DNA throughout the gradient for both the +As/-P and –As/+P cells (Fig. 4A and Table S2).

Likewise, no arsenate-conjugated mono- or dinucleotides were detected by exact mass (Figs. 4B and 4D). Although retention time and ionization efficiency could not be validated using standards for these molecules, their behavior, if the molecules were stable, would be expected to resemble their phosphorylated analogs sufficiently to allow detection. Finally, an enrichment of deoxynucleosides per ng DNA obtained from GFAJ-1 grown in the +As/-P condition, relative to either –As/+P or –As/-P conditions, could indicate nicked DNA resulting from arsenate-ester hydrolysis. However, we did not detect any enrichment despite detecting deoxyadenosine, deoxyguanosine, deoxycytidine, and thymidine (Fig. S1 and Table S2). Thus, while we detected arsenate associated with GFAJ-1 DNA, we found no evidence for arsenate bound sufficiently tightly to resist washing with water, nor able to co-migrate with the DNA in a CsCl gradient. Differences in DNA purity can readily explain the conflict of these results with Wolfe-Simon *et al.*'s claim that GFAJ-1 uses arsenate to replace scarce phosphate in its DNA.

Our LC-MS analyses rule out incorporation of arsenic in DNA at the ~ 0.1% level, and a much lower limit is suggested by our gel analysis of DNA integrity. Given the chemical similarity of arsenate to phosphate, it is likely that GFAJ-1 may sometimes assimilate arsenate into some small molecules in place of phosphate, such as sugar phosphates or nucleotides. Although the ability to tolerate or correct very low level incorporation of arsenic into DNA could be a contributor to the arsenate resistance of GFAJ-1, such low level incorporation would not be a biologically functional substitute for phosphate, and thus would have no significant impact on the organism's requirements for phosphate.

From a broader perspective, GFAJ-1 cells growing in Mono Lake face the challenge of discriminating an essential salt ($PO_4$, 400 μM) from a highly abundant but toxic chemical mimic ($AsO_4$, 200 μM). Similar salt management challenges are encountered by many other microorganisms, for instance those growing in environments with scarce potassium and plentiful ammonia (*8*). Organisms typically adapt to such conditions not by incorporating the mimic in place of the essential salt but by enriching for the salt at multiple stages, from preferential membrane transport to the selectivity of metabolic enzymes. The end result is that the fundamental biopolymers conserved across all forms of life remain, in terms of chemical backbone, invariant.

**Supplementary Materials:**

www.sciencemag.org

Materials and Methods

Fig. S1

References (9-11)

Tables S1 to S2

## Acknowledgements:

MLR is supported by a Graduate Research Fellowship from the National Science Foundation. JDR is supported by the CAREER Award from the National Science Foundation. LK is an Investigator of the Howard Hughes Medical Institute and a James S. McDonnell Foundation Centennial Fellow. RJR thanks the Canadian Institutes of Health Research for funding, Jodi Blum and Ron Oremland for providing strain GFAJ-1, Maryam Khoshnoodi for the trace element mix, the Charles Thompson lab for use of their BioScreen Analyzer, and Simon Silver and Christopher Rensing for helpful discussions. We also thank the ICP Laboratory in the Department of Geosciences at Princeton University for assistance with ICP-MS analysis.

**Figure Captions:**

**Fig. 1. Growth curves of GFAJ-1 in AML60 medium supplemented with different concentrations of phosphate**. Each line is the mean of 10 replicate 300 µl cultures in wells of a Bioscreen C Growth Analyzer. The phosphate additions used to replicate the '-P' and '+P' conditions of (*1*) are indicated.

**Fig. 2. Integrity of GFAJ-1 chromosomal DNA after long-term storage**. Lanes: 1 and 7: HindIII digest of lambda DNA; 2: *H. influenzae* chromosomal DNA; 3-6: GFAJ-1 chromosomal DNA grown in the specified combinations of As and P (-As: no arsenate; +As, 40 mM arsenate; -P, 3 µM added phosphate, +P, 1500 µM added phosphate). Panel **A.** ~100 ng of GFAJ-1 DNA immediately after purification. Panel **B.** 200 ng of the same DNAs after 2 months storage in Tris EDTA at 4°C. Panel **C.** The same DNAs as **B**, but 800 ng/lane and after 10 min at 95 °C.

**Fig. 3. LC-MS analysis of arsenate in purified and CsCl fractioned DNA from arsenate-grown GFAJ-1 cells.** Representative extracted ion chromatograms for arsenate (mass-to-charge ratio, *m/z* = 140.9174 +/- 3 ppm) are shown as the chromatographic retention time in minutes plotted against intensity in ion counts. Sample identity is indicated to the right, along the axis extending into the page. DNA from arsenate-grown GFAJ-1 cells (+As/-P undigested gDNA) was analyzed by LC-MS at a 1:10 dilution, as were the water wash (+As/-P wash of gDNA), the same DNA following washing and enzymatic digestion (+As/-P washed, digested DNA), and finally, fractions of the same DNA after a CsCl gradient purification and digestion (+As/-P CsCl Fractions #1 - #8, with DNA concentrating in Fractions #6, #7, and #8). Potassium arsenate standards (Std 1.7e-6 to 1.7e-8 [M]) and a water blank were also analyzed. One of four representative experiments is shown.

**Fig. 4. LC-MS analysis of deoxynucleotides from purified and CsCl fractioned DNA from**

**arsenate-grown GFAJ-1 cells.** Representative extracted ion chromatograms are shown as the chromatographic retention time in minutes plotted against intensity in ion counts. One of four representative experiments is shown.

**A. and B.** Extracted ion chromatograms for **A.** deoxyadenosine-phosphate (dAMP; *m/z* = 330.0609 +/- 5 ppm) and **B.** its arsenate analog deoxyadenosine-arsenate (dAMA; *m/z* = 374.0087 +/- 5 ppm). DNA from arsenate grown GFAJ-1 cells (+As/-P washed, digested gDNA) was washed, digested, and analyzed by LC-MS, as was the same DNA following a CsCl gradient purification and digestion (+As/-P CsCl Fractions #1 - #8). To keep the peak on scale, the signal for +As/-P washed, digested gDNA has been multiplied by 0.5. This observed large peak matches the known retention time of dAMP.

**C. and D.** Extracted ion chromatograms for **C.** the dideoxynucleotide deoxyadenosine-phosphate (dAMP-dAMP; *m/z* = 643.1185 +/- 5 ppm) and **D.** its mono-arsenate analog deoxyadenosine-arsenate-deoxyadenosine-phosphate (dAMA-dAMP; *m/z* = 687.0663 +/- 5 ppm). DNA from arsenate grown GFAJ-1 cells (+As/-P washed, digested gDNA) was washed, digested, and analyzed by LC-MS, as was the same DNA following a CsCl gradient purification and digestion (+As/-P CsCl Fractions #1 - #8). Partially digested -As/+P DNA shows a large peak at the exact mass of dAMP-dAMP.

**Table 1.** DNA, arsenate and nucleotide content of samples measured by absorbance and LC-MS.

| Sample | $A_{260}$ (DNA) | Compound | | | | |
|---|---|---|---|---|---|---|
| | | Arsenate | dAMP | dAMA | dAMP-dAMP | dAMA-dAMP |
| | AU (µg) | Peak Area, ion counts | | | | |
| **Digested CsCl Fractions: (increasing density)** | | | | | | |
| #1 (top) | 0.03 (0) | 0 | 0 | 154 | 0 | 0 |
| #2 | 0.01 (0) | 0 | 0 | 0 | 0 | 0 |
| #3 | 0.02 (0) | 226 | 0 | 0 | 0 | 0 |
| #4 | 0.01 (0) | 0 | 160 | 0 | 0 | 0 |
| #5 | 0.01 (0) | 0 | 0 | 0 | 0 | 0 |
| #6 | 0.82 (4.5) | 157 | 39,000 | 0 | 0 | 0 |
| #7 | 1.12 (6.7) | 373 | 52,000 | 0 | 0 | 0 |
| #8 (bottom) | 0.44 (1.9) | 300 | 3,700 | 0 | 0 | 197 |
| Water blank | 0 (0) | 329 | 0 | 0 | 0 | 0 |
| -As/+P partial digest | (3.3) | 515 | 0 | 0 | 21,000 | 210 |
| +As/-P washed, digested DNA | (1.7) | 2625 | 186,457 | 0 | 241 | 202 |
| +As/-P whole DNA (1:10 dil.) | (1.7) | 2794 | 562 | 0 | 781 | 0 |
| +As/-P wash of gDNA (300µL) | 0 (0) | 9545 | 182 | 207 | 0 | 221 |
| **Arsenate Standards [molar]** | | | | | | |
| $1.66 \times 10^{-8}$ | | 329 | | | | |
| $1.66 \times 10^{-7}$ | | 1959 | | | | |
| $1.66 \times 10^{-5}$ | | 59,925 | | | | |
| *Expected if DNA As:P = 0.04* | (6.7) | ~122,000 | | *>0* | | *>0* |

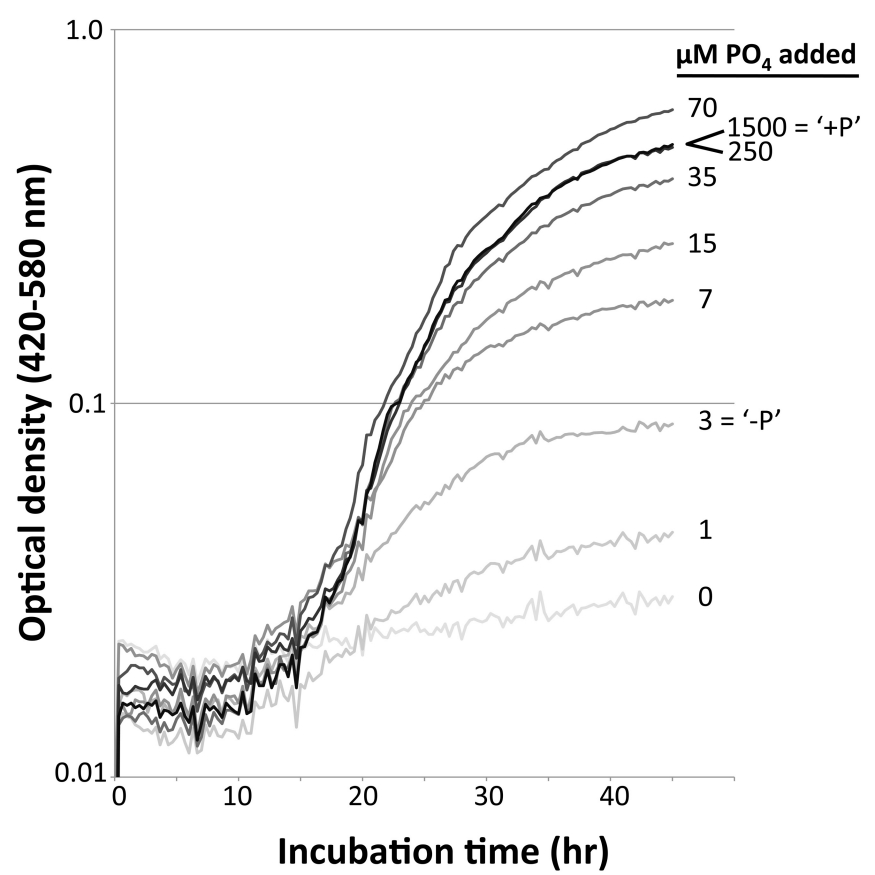

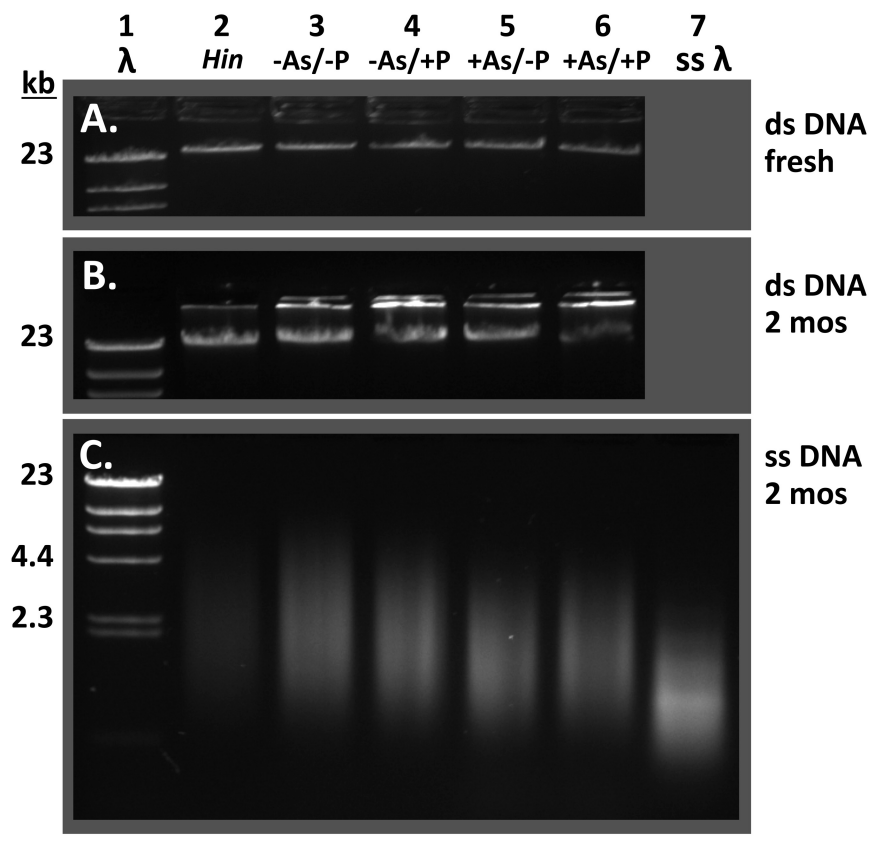

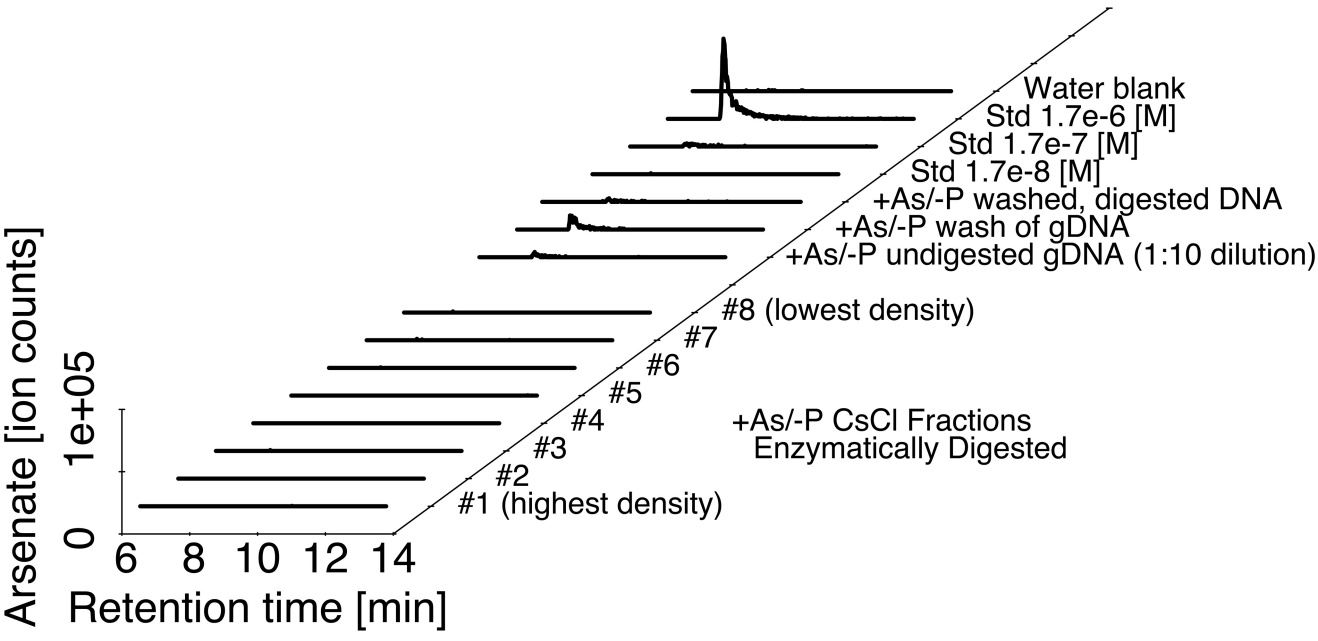

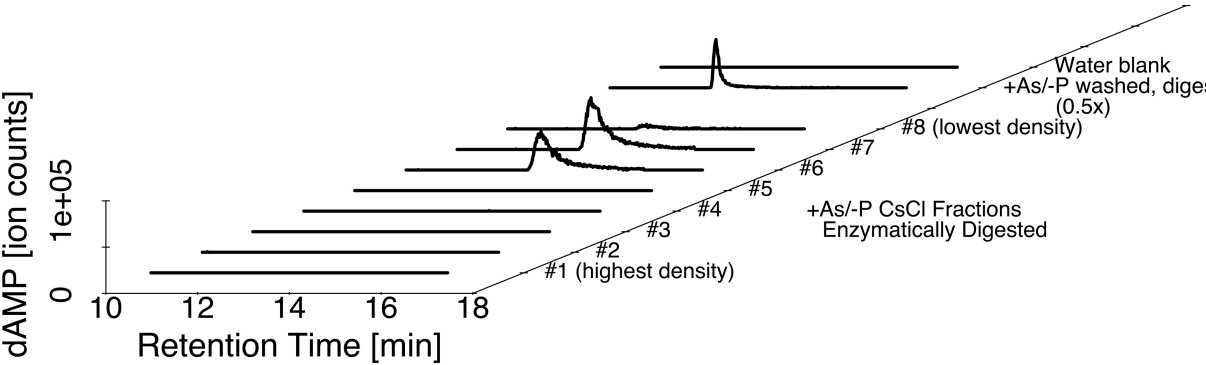

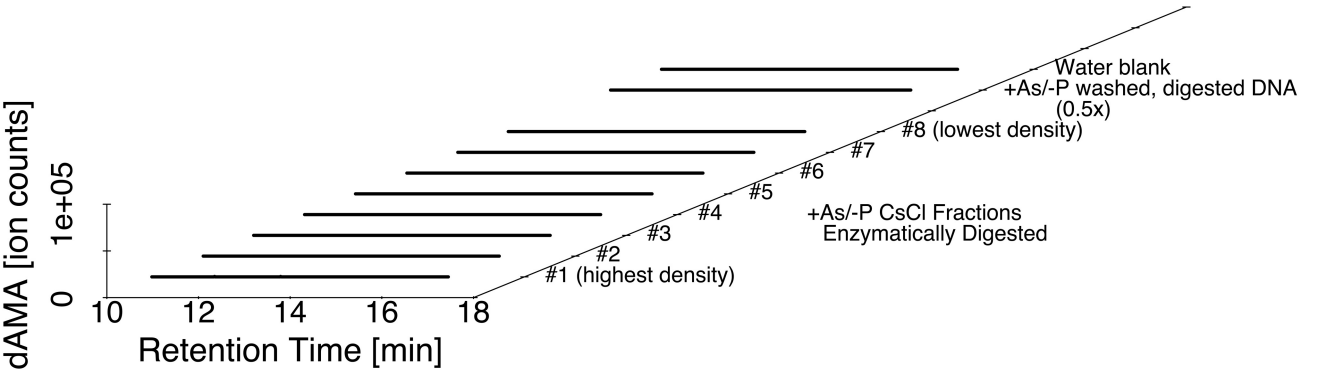

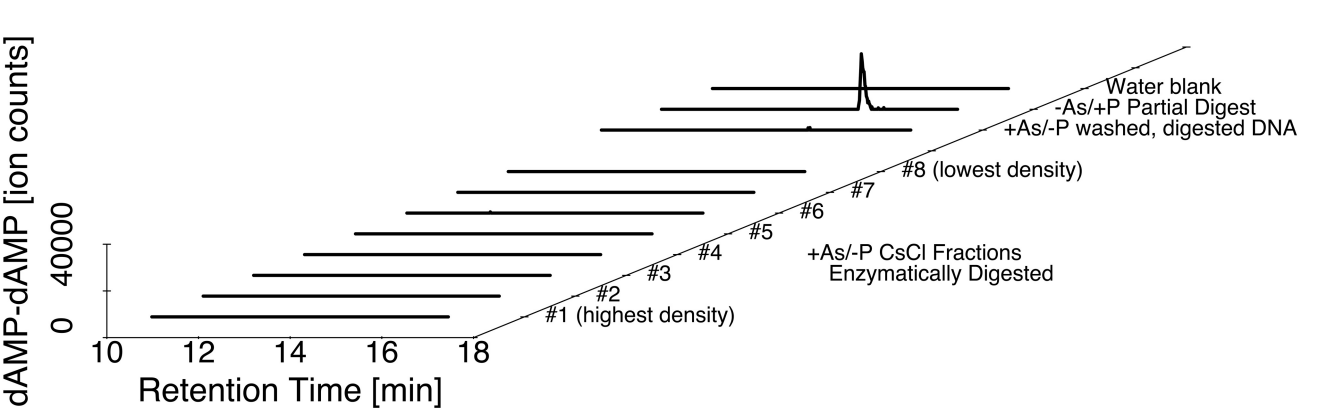

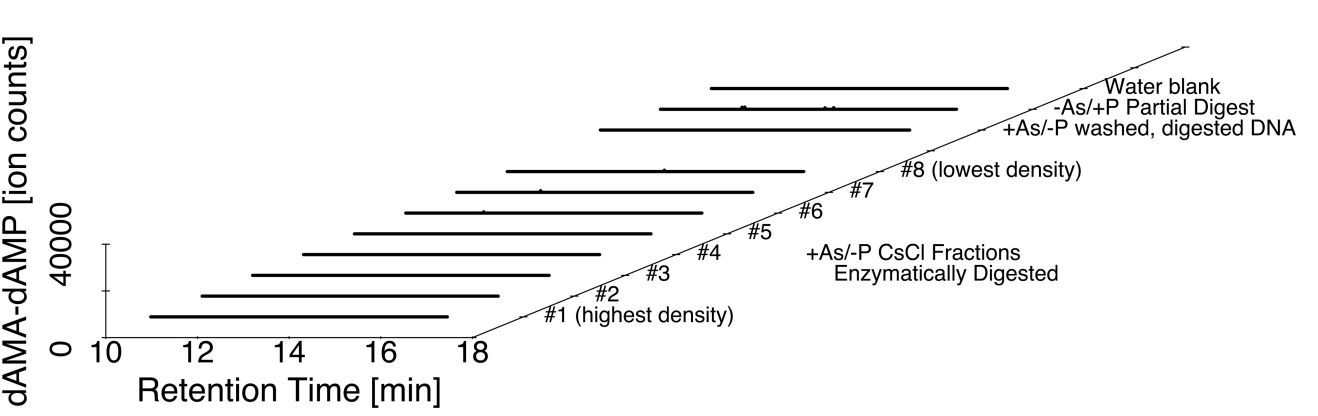

# Supplementary Materials for

**Absence of detectable arsenate in DNA from arsenate-grown GFAJ-1 cells.**

Marshall Louis Reaves[1,2], Sunita Sinha[3], Joshua D. Rabinowitz[1,4], Leonid Kruglyak[1,5,6], and Rosemary J. Redfield[3†].

correspondence to: redfield@zoology.ubc.ca

**This PDF file includes:**

## Materials and Methods

Culture methods: Reagent-grade chemicals and glass-distilled water were used throughout. Basal culture medium was AML60 buffered to pH 9.8 with carbonate rather than phosphate (1 M NaCl, 0.2 mM MgSO4, 100 mM Na2CO3, 50 mM NaHCO3, 0.8 mM (NH4)2 $SO_4$ and 10 mM glucose) and supplemented with vitamins and trace elements as specified by Wolfe-Simon *et al*. (*1*). For growth of GFAJ-1, this was further supplemented with 10 mM KCl and 1 mM sodium glutamate. $Na_3PO_4$ and $Na_3AsO_4$ (Sigma) were added as required. Final cell densities (as cfu/ml) were determined by plating diluted cultures on agar plates containing AML60 medium without glutamate but with 1500 µM phosphate and 5 mg/ml each of tryptone and yeast extract. Liquid cultures used for DNA preparation and growth analysis were inoculated with cells that had been frozen after depletion of their phosphate reserves by pre-growth in medium supplemented with 3 µM phosphate and 40 mM arsenate (to replicate the standard growth condition used in (*1*)), followed by thorough washing to remove external arsenate. Liquid cultures were incubated at 28°C, usually in half-full screw-capped glass tubes or bottles with gentle rocking. Two replicate sets of growth curves were done using a BioScreen C Growth Analyzer (Oy Growth Curves Ab Ltd) with 300 µl of each medium in each of 10 replicate wells; broadband readings of optical density (420-580 nm) were taken every 20 min, with 15 sec of shaking before each reading. Our '-P' and '+P' growth conditions were AML60 supplemented with glutamate, KCl, and 3 µM or 1500 µM $PO_4$ respectively.

DNA purification: Cells were pelleted, resuspended in 50 mM Tris 10 mM EDTA, lysed with 1% SDS and extracted with phenol and phenol:chloroform. Ethanol was then added to 70% and the DNA was spooled onto a glass rod and rinsed with 70% ethanol. The air-dried DNA was resuspended in 10 mM Tris 1 mM EDTA pH 8.0 (TE) at a concentration of ~ 100 µg/ml, incubated with RNase A (100 µg/ml) and Proteinase K (1 mg/ml) for 30 min at 37 °C, extracted again with phenol and phenol chloroform, and again spooled from 70% ethanol and rinsed. The DNA was air-dried, resuspended in TE, and stored at 4 °C.

CsCl gradients: Between 6300 and 33000 ng of purified DNA in 100 µL of TE was added to 900 µL of an aqueous solution of cesium chloride (Sigma) with density of 1.8 g/mL. After gentle mixing, this was pipetted into a polyallomer tube (Beckman Coulter), topped with approximately 3 mL of mineral oil and centrifuged in an SW-60 Ti rotor at 30,000 rpm for 18 hours at 20°C followed by unaided deceleration. Immediately following centrifugation, fractions were collected from bottom (most dense) to top (least dense) using a water-based backpressure device. DNA content of fractions was measured using a NanoDrop instrument (ThermoFisher).

DNA cleanup and limit digestion: Prior to digestion, DNA was washed with 350 µL DNAse-free water using 3 kDa centrifugal membranes (Amicon). A two-enzyme limit digestion protocol was used based on previous studies (*9*): ~40 µL was treated with 0.2 U of P1 nuclease (US Biological, Swampscott, MA) for 2 hours at 50°C followed by the addition of 50 µL 30 mM sodium acetate (pH 8.1) and 1 U snake venom exonuclease (US Biological) and incubation for 18 hours at 37°C. Finally, protein was removed by filtration using 3 kDa centrifugal membranes.

LC-MS methods: Free arsenate and both arsenate and phosphate analogs of deoxynucleotides and dideoxynucleotides, were measured using a previously described LC-MS method (*10*). In this analysis samples are first chemically separated by hydrophobicity using liquid chromatography (LC), and then the separated species are ionized by electrospray and the quantity of ions of different exact masses are measured using mass spectrometry (MS) with 100,000 resolving power (m/Δm). The combination of these two measurements, retention time and intensity at a specific m/z, typically provides unambiguous identification of small molecule metabolites. Analysis of standards of known concentration allows the amounts of specific metabolites to be quantitated. The results for a 5 parts per million (ppm) mass-to-charge window around each metabolite of interest were examined separately, as plots showing ion intensity as a function of retention time (extracted ion chromatograms). Briefly, eluant from an Accela U-HPLC system with quaternary pumps was ionized by electrospray into an Exactive orbitrap mass spectrometer operating in negative mode (both from ThermoFisher Scientific, San Jose, CA) under control of Xcalibur 2.1 software. Liquid chromatography separation was achieved on a Synergy Hydro-RP column (*10*) (100 mm × 2 mm, 2.5 μm particle size, Phenomenex, Torrance, CA), using reversed- phase chromatography with the ion pairing agent tributylamine in the aqueous mobile phase. The flow rate was 200 μL/min; solvent A 97:3 water/methanol with 10 mM tributylamine and 15 mM acetic acid; solvent B methanol. The gradient was 0 min, 0% B; 2.5 min, 0% B; 5 min, 20% B; 7.5 min, 20% B; 13 min, 55% B; 15.5 min, 95% B; 18.5 min, 95% B; 19 min, 0% B; 25 min, 0% B. Other LC parameters were: autosampler temperature 4 °C, injection volume 10 μL, and column temperature 25 °C, maintained by a Keystone hot pocket column heater (ThermoFisher). Mass calibration was performed weekly using polytyrosine-1,3,6 standards (ThermoFisher). Analysis of LC-MS data was performed using the MAVEN software program with mass domain resolution of +/- 5 ppm (*11*).

<u>ICP-MS methods:</u> Media was analyzed for phosphate concentrations on a Thermo Element 2 single collector ICP-MS in the ICP Laboratory in the Department of Geosciences at Princeton University. Solutions were measured under both medium and high resolution with identical phosphate concentrations obtained by both measurements. Correlation coefficients for the calibration curves were 0.995. External reproducibility of repeat measurements was 6-8%.

**Fig. S1.**

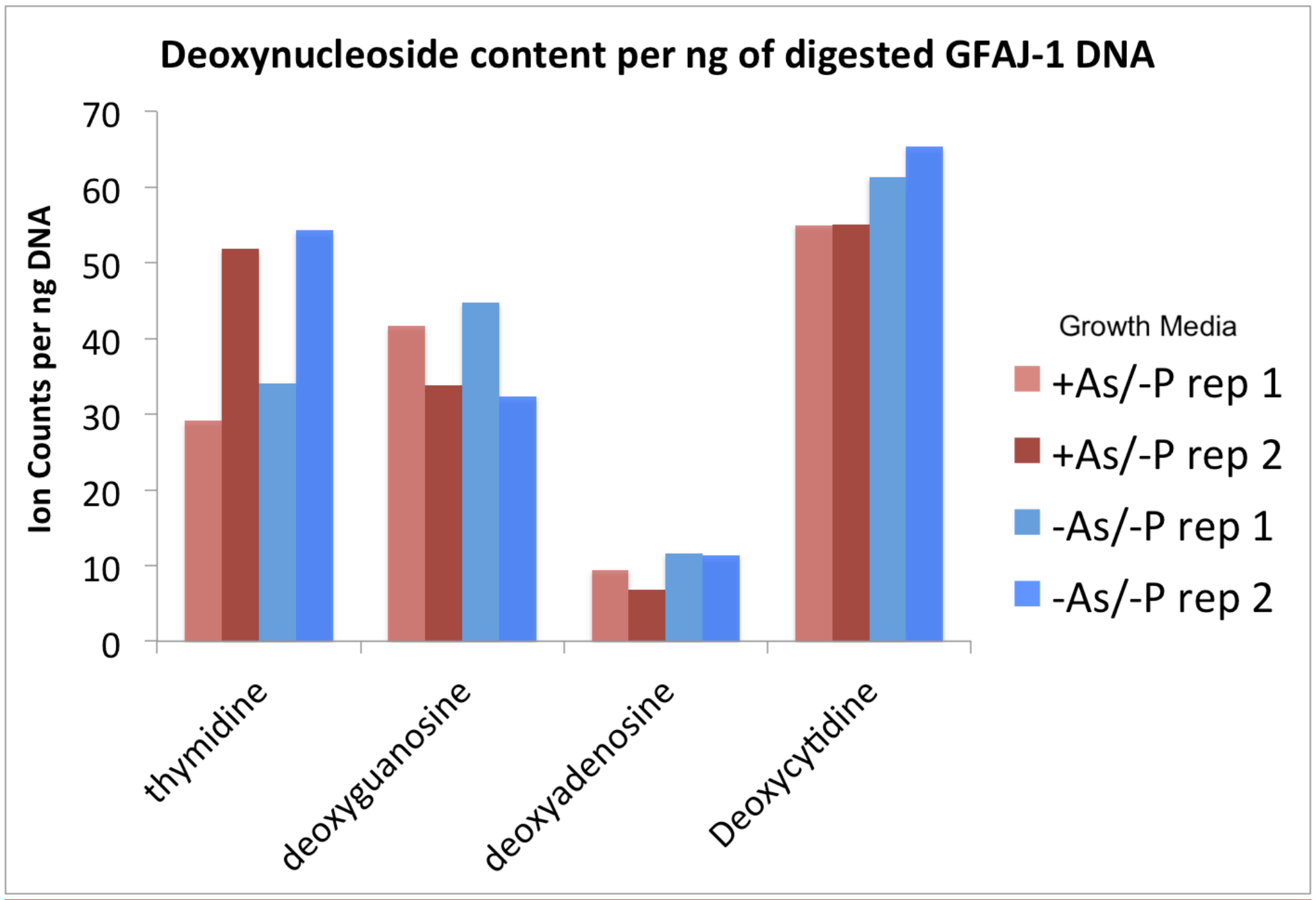

Digested GFAJ-1 +As/-P DNA is not enriched for deoxynucleosides. 6300 ng of DNA from conditions with (+As/-P) and without (-As/+P) added arsenate were washed with water and enzymatically digested. Ion counts per ng DNA for each deoxynucleoside measured by LC-MS are shown for two independent experiments.

**Table S1.**

Details of arsenate and DNA measurements using LC-MS.

| | Analysis Date | CsCl gradient fraction # | A260 | A280 | Vol. [µL] | Arsenate peak area top | Arsenate median peak m/z | Arsenate retention time [min] | DNA source |
|---|---|---|---|---|---|---|---|---|---|
| +As/-P CsCl purified, digested | 01/15/12 | 1 | 0.031 | 0.002 | 110 | 0 | | | Batch #1 |
| +As/-P CsCl purified, digested | 01/15/12 | 2 | 0.013 | -0.006 | 80 | 0 | | | Batch #1 |
| +As/-P CsCl purified, digested | 01/15/12 | 3 | 0.021 | 0.012 | 100 | 227 | 140.9168 | 7.04 | Batch #1 |
| +As/-P CsCl purified, digested | 01/15/12 | 4 | 0.007 | 0.005 | 120 | 0 | | | Batch #1 |
| +As/-P CsCl purified, digested | 01/15/12 | 5 | 0.012 | 0.012 | 140 | 0 | | | Batch #1 |
| +As/-P CsCl purified, digested | 01/15/12 | 6 | 0.817 | 0.418 | 110 | 157 | 140.9168 | 6.96 | Batch #1 |
| +As/-P CsCl purified, digested | 01/15/12 | 7 | 1.12 | 0.579 | 120 | 216 | 140.9169 | 7.07 | Batch #1 |
| +As/-P CsCl purified, digested | 01/15/12 | 8 | 0.444 | 0.251 | 90 | 298 | 140.9167 | 6.87 | Batch #1 |
| | | | | | | | | | |
| +As/-P CsCl purified, digested | 12/30/11 | 1 | 0.157 | 0.065 | 90 | 0 | | | Batch #1 |
| +As/-P CsCl purified, digested | 12/30/11 | 2 | 0.095 | 0.019 | 180 | 0 | | | Batch #1 |
| +As/-P CsCl purified, digested | 12/30/11 | 3 | 0.075 | 0.03 | 80 | 0 | | | Batch #1 |
| +As/-P CsCl purified, digested | 12/30/11 | 4 | 0.049 | -0.004 | 70 | 0 | | | Batch #1 |
| +As/-P CsCl purified, digested | 12/30/11 | 5 | 0.053 | 0.015 | 70 | 0 | | | Batch #1 |
| +As/-P CsCl purified, digested | 12/30/11 | 6 | 0.07 | 0.024 | 70 | 0 | | | Batch #1 |
| +As/-P CsCl purified, digested | 12/30/11 | 7 | 1.33 | 0.712 | 50 | 0 | | | Batch #1 |
| +As/-P CsCl purified, | 12/30/11 | 8 | 1.417 | 0.765 | 80 | 178 | 140.9174 | 6.69 | Batch #1 |
| digested | | | | | | | | | |
| +As/-P CsCl purified, digested | 12/30/11 | 9 | 0.686 | 0.365 | 80 | 0 | | | Batch #1 |
| +As/-P CsCl purified, digested | 12/30/11 | 10 | 0.101 | 0.045 | 100 | 0 | | | Batch #1 |
| -As/-P CsCl purified, digested | 01/11/12 | 1 | 0.161 | 0.125 | 10 | 179 | 140.9169 | 6.96 | Batch #1 |
| -As/-P CsCl purified, digested | 01/11/12 | 2 | 0.119 | 0.076 | 80 | 192 | 140.9169 | 8.91 | Batch #1 |
| -As/-P CsCl purified, digested | 01/11/12 | 3 | 0.096 | 0.057 | 100 | 0 | | | Batch #1 |
| -As/-P CsCl purified, digested | 01/11/12 | 4 | 0.1 | 0.077 | 90 | 212 | 140.9168 | 8.10 | Batch #1 |
| -As/-P CsCl purified, digested | 01/11/12 | 5 | 0.078 | 0.031 | 80 | 0 | | | Batch #1 |
| -As/-P CsCl purified, digested | 01/11/12 | 6 | 0.076 | 0.04 | 20 | 0 | | | Batch #1 |
| -As/-P CsCl purified, digested | 01/11/12 | 7 | 0.08 | 0.044 | 80 | 0 | | | Batch #1 |
| -As/-P CsCl purified, digested | 01/11/12 | 8 | 0.095 | 0.058 | 50 | 192 | 140.9172 | 7.64 | Batch #1 |
| -As/-P CsCl purified, digested | 01/11/12 | 9 | 0.346 | 0.196 | 70 | 0 | | | Batch #1 |
| -As/-P CsCl purified, digested | 01/11/12 | 10 | 2.983 | 1.618 | 70 | 0 | | | Batch #1 |
| -As/-P CsCl purified, digested | 01/11/12 | 11 | 1.692 | 0.913 | 200 | 0 | | | Batch #1 |
| -As/-P CsCl purified, digested | 01/11/12 | 12 | 0.333 | 0.18 | 70 | 0 | | | Batch #1 |
| -As/-P CsCl purified, digested | 01/11/12 | 13 | 0.615 | 0.335 | 70 | 0 | | | Batch #1 |
| -As/-P CsCl purified, digested | 12/30/11 | 1 | 0.028 | 0.006 | 100 | 0 | | | Batch #1 |
| -As/-P CsCl purified, digested | 12/30/11 | 2 | 0.04 | 0.005 | 10 | 0 | | | Batch #1 |
| -As/-P CsCl | 12/30/11 | 3 | 0.024 | 0.002 | 30 | 0 | | | Batch |
| purified, digested | | | | | | | | | #1 |
| -As/-P CsCl purified, digested | 12/30/11 | 4 | 0.028 | 0.012 | 50 | 0 | | | Batch #1 |
| -As/-P CsCl purified, digested | 12/30/11 | 5 | -0.001 | -0.034 | 20 | 0 | | | Batch #1 |
| -As/-P CsCl purified, digested | 12/30/11 | 6 | 0.03 | 0.015 | 40 | 0 | | | Batch #1 |
| -As/-P CsCl purified, digested | 12/30/11 | 7 | 0.017 | -0.001 | 20 | 0 | | | Batch #1 |
| -As/-P CsCl purified, digested | 12/30/11 | 8 | 0.034 | 0.02 | 110 | 205 | 140.9175 | 6.96 | Batch #1 |
| -As/-P CsCl purified, digested | 12/30/11 | 9 | 0.016 | 0.002 | 110 | 0 | | | Batch #1 |
| -As/-P CsCl purified, digested | 12/30/11 | 10 | 0.016 | -0.001 | 20 | 0 | | | Batch #1 |
| -As/-P CsCl purified, digested | 12/30/11 | 11 | 1.138 | 0.611 | 90 | 0 | | | Batch #1 |
| -As/-P CsCl purified, digested | 12/30/11 | 12 | 0.559 | 0.28 | 280 | 0 | | | Batch #1 |
| | | | | | | | | | |
| +As/-P washed digested DNA | 01/15/12 | | | | | 2409 | 140.9175 | 7.03 | Batch #1 |
| +As/-P whole gDNA (1:10 dilution) | 01/15/12 | | | | | 2795 | 140.9175 | 7.06 | Batch #1 |
| -As/+P washed digested DNA | 01/15/12 | | | | | 310 | 140.9175 | 7.55 | Batch #1 |
| -As/-P whole gDNA (1:10 dilution) | 01/15/12 | | | | | 0 | | | Batch #1 |
| | | | | | | | | | |
| +As/-P Wash #1 | 01/15/12 | | | | 350 | 180766 | 140.9175 | 6.99 | Batch #1 |
| +As/-P Wash #2 | 01/15/12 | | | | 350 | 3193 | 140.9174 | 7.24 | Batch #1 |
| +As/-P Wash #3 | 01/15/12 | | | | 350 | 1651 | 140.9174 | 7.10 | Batch #1 |
| | | | | | | | | | |
| -As/-P Wash #1 | 01/15/12 | | | | 350 | 0 | | | Batch #1 |
| -As/-P Wash #2 | 01/15/12 | | | | 350 | 0 | | | Batch #1 |
| -As/-P Wash | 01/15/12 | | | | 350 | 0 | | | Batch |
| #3 | | | | | | | | | #1 |
| **Arsenate standard concentration [M]** | | | | | | | | | |
| 1.66E-04 | 01/11/12 | | | | | 2454503 | 140.917 | 7.03 | |
| 1.66E-05 | 01/11/12 | | | | | 544647 | 140.9169 | 7.01 | |
| 1.66E-06 | 01/11/12 | | | | | 25281 | 140.9168 | 7.03 | |
| **Arsenate standard concentration [M]** | | | | | | | | | |
| 1.66E-05 | 01/15/12 | | | | | 876672 | 140.9176 | 7.06 | |
| 1.66E-06 | 01/15/12 | | | | | 59994 | 140.9176 | 7.07 | |
| 1.66E-07 | 01/15/12 | | | | | 2516 | 140.9176 | 7.04 | |
| 1.66E-07 | 01/15/12 | | | | | 1959 | 140.9175 | 7.10 | |
| 1.66E-08 | 01/15/12 | | | | | 344 | 140.9175 | 7.24 | |
| 1.66E-08 | 01/15/12 | | | | | 243 | 140.9174 | 7.15 | |
| 1.66E-09 | 01/15/12 | | | | | 381 | 140.9175 | 6.98 | |
| water blank | 01/10/12 | | | | | 192 | 140.9181 | 7.50 | |
| water blank | 01/11/12 | | | | | 188 | 140.9169 | 7.10 | |
| water blank | 01/11/12 | | | | | 268 | 140.917 | 7.60 | |
| water blank | 01/15/12 | | | | | 980 | 140.9174 | 7.50 | |
| water blank | 01/15/12 | | | | | 286 | 140.9175 | 7.44 | |
| -As/-P CsCl purified, digested | 01/20/12 | 1 | 0.021 | 0.038 | 75 | 305 | 140.9171 | 6.99 | Batch #2 |
| -As/-P CsCl purified, digested | 01/20/12 | 2 | 0.04 | 0.053 | 40 | 164 | 140.9172 | 7.67 | Batch #2 |
| -As/-P CsCl purified, digested | 01/20/12 | 3 | -0.048 | -0.102 | 220 | 517 | 140.9171 | 7.10 | Batch #2 |
| -As/-P CsCl purified, digested | 01/20/12 | 4 | 0.383 | 0.236 | 30 | 522 | 140.9171 | 7.07 | Batch #2 |
| -As/-P CsCl purified, digested | 01/20/12 | 5 | 0.547 | 0.294 | 165 | 337 | 140.9172 | 7.07 | Batch #2 |
| -As/-P CsCl purified, digested | 01/20/12 | 6 | 0.031 | 0.024 | 85 | 197 | 140.9171 | 7.21 | Batch #2 |
| -As/-P CsCl purified, digested | 01/20/12 | 7 | 0.004 | -0.004 | 130 | 219 | 140.9172 | 7.13 | Batch #2 |
| -As/-P CsCl | 01/20/12 | 8 | 0.028 | 0.056 | 70 | 229 | 140.9171 | 7.19 | Batch |
| purified, digested | | | | | | | | | #2 |
| -As/+P CsCl purified, digested | 01/20/12 | 1 | 0.043 | 0.056 | 40 | 598 | 140.9173 | 7.13 | Batch #2 |
| -As/+P CsCl purified, digested | 01/20/12 | 2 | 0.025 | 0.022 | 30 | 386 | 140.9172 | 7.28 | Batch #2 |
| -As/+P CsCl purified, digested | 01/20/12 | 3 | 0.03 | 0.041 | 40 | 379 | 140.9171 | 7.18 | Batch #2 |
| -As/+P CsCl purified, digested | 01/20/12 | 4 | 0.044 | 0.039 | 40 | 319 | 140.9171 | 7.21 | Batch #2 |
| -As/+P CsCl purified, digested | 01/20/12 | 5 | 0.028 | 0.04 | 50 | 258 | 140.9171 | 7.12 | Batch #2 |
| -As/+P CsCl purified, digested | 01/20/12 | 6 | 0.04 | 0.037 | 105 | 192 | 140.9171 | 7.03 | Batch #2 |
| -As/+P CsCl purified, digested | 01/20/12 | 7 | 0.354 | 0.203 | 90 | 661 | 140.9174 | 7.17 | Batch #2 |
| -As/+P CsCl purified, digested | 01/20/12 | 8 | 0.382 | 0.221 | 70 | 389 | 140.9172 | 7.23 | Batch #2 |
| -As/+P CsCl purified, digested | 01/20/12 | 9 | 0.276 | 0.208 | 140 | 203 | 140.9171 | 7.14 | Batch #2 |
| -As/+P CsCl purified, digested | 01/20/12 | 10 | 0.013 | 0.034 | 45 | 216 | 140.9172 | 7.36 | Batch #2 |
| -As/+P CsCl purified, digested | 01/20/12 | 11 | 0.01 | 0.028 | 60 | 241 | 140.9172 | 7.13 | Batch #2 |
| +As/-P CsCl purified, digested | 01/20/12 | 1 | 0.122 | 0.081 | 60 | 0 | | | Batch #2 |
| +As/-P CsCl purified, digested | 01/20/12 | 2 | 0.054 | 0.037 | 75 | 542 | 140.9173 | 6.96 | Batch #2 |
| +As/-P CsCl purified, digested | 01/20/12 | 3 | 0.034 | 0.018 | 70 | 342 | 140.9171 | 7.10 | Batch #2 |
| +As/-P CsCl purified, digested | 01/20/12 | 4 | 0.051 | 0.071 | 70 | 240 | 140.9173 | 6.98 | Batch #2 |
| +As/-P CsCl purified, digested | 01/20/12 | 5 | 0.03 | 0.023 | 50 | 503 | 140.9171 | 7.13 | Batch #2 |
| +As/-P CsCl purified, digested | 01/20/12 | 6 | 0.17 | 0.094 | 35 | 272 | 140.9172 | 7.32 | Batch #2 |
| +As/-P CsCl purified, digested | 01/20/12 | 7 | 0.364 | 0.234 | 35 | 214 | 140.9171 | 7.20 | Batch #2 |
| +As/-P CsCl purified, digested | 01/20/12 | 8 | 0.399 | 0.233 | 30 | 201 | 140.9174 | 7.39 | Batch #2 |
| +As/-P CsCl purified, digested | 01/20/12 | 9 | 0.608 | 0.355 | 50 | 283 | 140.9171 | 7.26 | Batch #2 |
| +As/-P CsCl purified, digested | 01/20/12 | 10 | 0.419 | 0.236 | 45 | 243 | 140.9173 | 7.10 | Batch #2 |
| +As/-P CsCl purified, digested | 01/20/12 | 11 | 0.411 | 0.242 | 45 | 178 | 140.9171 | 7.33 | Batch #2 |
| +As/-P CsCl purified, digested | 01/20/12 | 12 | 0.309 | 0.194 | 45 | 193 | 140.9169 | 8.75 | Batch #2 |
| +As/-P CsCl purified, digested | 01/20/12 | 13 | 0.023 | 0.028 | 55 | 224 | 140.9172 | 7.20 | Batch #2 |
| +As/-P CsCl purified, digested | 01/20/12 | 14 | 0.025 | 0.018 | 55 | 277 | 140.917 | 7.03 | Batch #2 |
| +As/-P CsCl purified, digested | 01/20/12 | 15 | 0.011 | 0.028 | 75 | 0 | | | Batch #2 |
| +As/-P CsCl purified, digested | 01/20/12 | 16 | 0.006 | 0.016 | 40 | 0 | | | Batch #2 |
| +As/-P CsCl purified, digested | 01/20/12 | 17 | 0.016 | 0.031 | 50 | 0 | | | Batch #2 |
| | | | | | | | | | |
| +As/-P CsCl purified, digested | 01/20/12 | 1 | 0.101 | 0.075 | 80 | 214 | 140.9172 | 6.97 | Batch #2 |
| +As/-P CsCl purified, digested | 01/20/12 | 2 | 0.018 | 0.028 | 140 | 289 | 140.9172 | 7.07 | Batch #2 |
| +As/-P CsCl purified, digested | 01/20/12 | 3 | -0.001 | 0.004 | 135 | 669 | 140.9172 | 7.15 | Batch #2 |
| +As/-P CsCl purified, digested | 01/20/12 | 4 | 0.355 | 0.199 | 60 | 299 | 140.9173 | 7.07 | Batch #2 |
| +As/-P CsCl purified, digested | 01/20/12 | 5 | 0.763 | 0.426 | 40 | 439 | 140.9172 | 7.19 | Batch #2 |
| +As/-P CsCl purified, digested | 01/20/12 | 6 | 0.26 | 0.163 | 65 | 269 | 140.9172 | 7.26 | Batch #2 |
| +As/-P CsCl purified, digested | 01/20/12 | 7 | 0.349 | 0.218 | 40 | 191 | 140.9174 | 7.12 | Batch #2 |
| +As/-P CsCl purified, digested | 01/20/12 | 8 | 0.021 | 0.03 | 160 | 189 | 140.9171 | 7.05 | Batch #2 |
| +As/-P CsCl purified, digested | 01/20/12 | 9 | 0.184 | 0.111 | 40 | 0 | | | Batch #2 |
| +As/-P CsCl purified, digested | 01/20/12 | 10 | 0.034 | 0.038 | 55 | 0 | | | Batch #2 |
| **Arsenate standard concentration [M]** | | | | | | | | | |
| 1.06E-08 | 01/20/12 | | | | | 757 | 140.9172 | 7.03 | |
| 1.06E-08 | 01/20/12 | | | | | 430 | 140.9172 | 7.13 | |
| 5.31E-08 | 01/20/12 | | | | | 191 | 140.9172 | 7.32 | |
| 5.31E-08 | 01/20/12 | | | | | 486 | 140.9171 | 7.09 | |
| 2.66E-07 | 01/20/12 | | | | | 632 | 140.9172 | 7.12 | |
| 2.66E-07 | 01/20/12 | | | | | 479 | 140.9171 | 7.21 | |
| 2.66E-07 | 01/20/12 | | | | | 430 | 140.9171 | 7.09 | |
| 1.33E-06 | 01/20/12 | | | | | 4907 | 140.9172 | 7.12 | |
| 1.33E-06 | 01/20/12 | | | | | 3486 | 140.9172 | 7.17 | |
| 6.64E-06 | 01/20/12 | | | | | 36873 | 140.9172 | 7.12 | |
| 6.64E-06 | 01/20/12 | | | | | 28012 | 140.9172 | 7.10 | |
| 3.32E-05 | 01/20/12 | | | | | 346429 | 140.9172 | 7.09 | |
| 3.32E-05 | 01/20/12 | | | | | 309439 | 140.9172 | 7.09 | |
| -As/-P washed digested DNA | 01/20/12 | | | | | 467 | 140.9171 | 6.93 | Batch #2 |
| +As/-P washed digested DNA | 01/20/12 | | | | | 2625 | 140.9172 | 7.12 | Batch #2 |
| -As/+P washed digested DNA | 01/20/12 | | | | | 892 | 140.9172 | 7.02 | Batch #2 |
| +As/-P whole gDNA (1 to 10 dilution) | 01/20/12 | | | | | 3984 | 140.9172 | 7.23 | Batch #2 |
| -As/+P whole gDNA (1 to 10 dilution) | 01/20/12 | | | | | 0 | | | Batch #2 |
| -As/-P whole gDNA (1 to 10 dilution) | 01/20/12 | | | | | 0 | | | Batch #2 |
| +As/-P Wash #1 | 01/20/12 | | | | 350 | 1481 | 140.9171 | 7.10 | Batch #2 |
| -As/-P Wash #1 | 01/20/12 | | | | 350 | 0 | | | Batch #2 |

| | | | | | | |
|---|---|---|---|---|---|---|
| -As/+P Wash #1 | 01/20/12 | 350 | 0 | | | Batch #2 |
| +As/-P Wash #2 | 01/20/12 | 350 | 222 | 140.9173 | 7.12 | Batch #2 |
| -As/-P Wash #2 | 01/20/12 | 350 | 0 | | | Batch #2 |
| -As/+P Wash #2 | 01/20/12 | 350 | 0 | | | Batch #2 |
| +As/-P Wash #3 | 01/20/12 | 350 | 0 | | | Batch #2 |
| -As/-P Wash #3 | 01/20/12 | 350 | 0 | | | Batch #2 |
| -As/+P Wash #3 | 01/20/12 | 350 | 0 | | | Batch #2 |
| water blank | 01/20/12 | | 329 | 140.9172 | 7.29 | |
| water blank | 01/20/12 | | 393 | 140.9172 | 7.60 | |

**Table S2.**

Details of nucleotide and nucleoside measurements using LC-MS.

| Compound | Formula | median m/z | retention time [min] | Notes |
|---|---|---|---|---|
| **Mononucleosides** | | | | |
| deoxycytidine-phosphate (dCMP) | C9H14N3O7P | 306.0502 | 10.86 | |
| deoxycytosine-arsenate (dCMA) | C9H14N3O7As | 349.9970 | 10.74 | |
| deoxyguanosine-phosphate (dGMP) | C10H14N5O7P | 346.0565 | 11.96 | Some samples have AMP contamination in this mass channel @ 11.3 min |
| deoxyguanosine-arsenate (dGMA) | C10H14N5O7As | 390.0030 | 10.52 | |
| thymidine-phosphate (dTMP) | C10H15N2O8P | 321.0500 | 11.71 | |
| deoxythiamine-arsenate (dTMA) | C10H15N2O8As | 364.9968 | 12.41 | Adduct peak from both in -As and +As conditions @ 7.1 min |
| deoxyadenosine-phosphate (dAMP) | C10H14N5O6P | 330.0615 | 12.86 | |
| deoxyadenosine-arsenate (dAMA) | C10H14N5O6As | 374.0093 | 12.60 | |
| **Dideoxynucleotides** | | | | |
| deoxy-PO4-PO4-AA | C20H26N10O11P2 | 643.1198 | 14.26 | |
| deoxy-PO4-PO4-AC | C19H26N8O12P2 | 619.1072 | 14.06 | |
| deoxy-PO4-PO4-AG | C20H26N10O12P2 | 659.1146 | 14.04 | |
| deoxy-PO4-PO4-AT | C20H27N7O13P2 | 634.1086 | 14.22 | |
| deoxy-PO4-PO4-CC | C18H26N6O13P2 | 595.0977 | 13.47 | |
| deoxy-PO4-PO4-CT | C19H27N5O14P2 | 610.0974 | 13.85 | |
| deoxy-PO4-PO4-GC | C19H26N8O13P2 | 635.1033 | 13.72 | |
| deoxy-PO4-PO4-GG | C20H26N10O13P2 | 675.1100 | 13.89 | |
| deoxy-PO4-PO4-GT | C20H27N7O14P2 | 650.1035 | 13.99 | |
| deoxy-PO4-PO4-TT | C20H28N4O15P2 | 625.0967 | 14.16 | |
| deoxy-PO4-AsO4-AA | C20H26N10O11AsP | 687.0665 | 13.54 | Compound appears in washes of both -As and +As DNA |
| deoxy-PO4-AsO4-AC | C19H26N8O12AsP | 663.0554 | 13.99 | Adduct peak from both in -As and +As conditions |
| deoxy-PO4-AsO4-AG | C20H26N10O12AsP | 703.0588 | 11.71 | |
| deoxy-PO4-AsO4-AT | C20H27N7O13AsP | 678.0537 | 12.96 | |
| deoxy-PO4-AsO4-CC | C18H26N6O13AsP | 639.0443 | 14.27 | |
| deoxy-PO4-AsO4-CT | C19H27N5O14AsP | 654.0422 | 13.68 | |
| deoxy-PO4-AsO4-GC | C19H26N8O13AsP | 679.0501 | 10.96 | |
| deoxy-PO4-AsO4-GG | C20H26N10O13AsP | 719.0547 | 12.37 | Unknown signal in both -As and +As samples |
| deoxy-PO4-AsO4-GT | C20H27N7O14AsP | 694.0516 | 11.64 | |
| deoxy-PO4-AsO4-TT | C20H28N4O15AsP | 669.0435 | 11.31 | Adduct of dTMP detected in -As samples |